# Optimal Solid Space Tower*

**Alexander Bolonkin**
C&R, 1310 Avenue R, #F-6, Brooklyn, NY 11229, USA
T/F 718-339-4563, aBolonkin@juno.com, http://Bolonkin.narod.ru

## Abstract

Theory and computations are provided for building of optimal (minimum weight) solid space towers (mast) up to one hundred kilometers in height. These towers can be used for tourism; scientific observation of space, observation of the Earth's surface, weather and upper atmosphere experiment, and for radio, television, and communication transmissions. These towers can also be used to launch spaceships and Earth satellites.

These macroprojects are not expensive. They require strong hard material (steel). Towers can be built using present technology. Towers can be used (for tourism, communication, etc.) during the construction process and provide self-financing for further construction. The tower design does not require human work at high altitudes; the tower is separated into sections; all construction can be done at the Earth's surface.

The transport system for a tower consists of a small engine (used only for friction compensation) located at the Earth's surface.

Problems involving security, control, repair, and stability of the proposed towers are addressed in other cited publications.

*Presented as paper to 45th AIAA Aerospace Science Meeting, 8 - 11 January 2007, Reno, Nevada, USA.
(see details in author's works: AIAA-2006-4235, AIAA-2006-7717).

## 1. Introduction

**1.1. Brief History.** The idea of building a tower high above the Earth into the heavens is very old [1]. The writings of Moses, in chapter 11 of his book *Genesis* refers to an early civilization that tried to build a tower to heaven out of brick and tar. This construction was called the Tower of Babel, and was reported to be located in Babylon in ancient Mesopotamia. Later in chapter 28, Jacob had a dream about a staircase or ladder built to heaven. This construction was called Jacob's Ladder. More contemporary writings on the subject date back to K.E. Tsiolkovski in his manuscript "Speculation about Earth and Sky and on Vesta," published in 1895 [2]. This idea inspired Sir Arthur Clarke to write his novel, *The Fountains of Paradise* [3], about a space tower (elevator) located on a fictionalized Sri Lanka, which brought the concept to the attention of the entire 20th Century world.

Today, the world's tallest construction is a television transmitting tower (mast) near Fargo, North Dakota, USA. It stands 629 m high and was built in 1963 for KTHI-TV. The CNN Tower in Toronto, Ontario, Canada is the world's tallest building. It is 553 m in height, was completed in1975, and has the world's highest observation deck at 447 m. The tower structure is concrete up to the observation deck level. Above is a steel structure supporting radio, television, and communication antennas. The total weight of the tower is 3,000,000 metric tons.

The Ostankin Tower in Moscow is 540 m in height and has an observation desk at 370 m. The world's tallest office building is the Petronas Towers in Kuala Lumpur, Malasia. The twin towers are 452 m in height. They are 10 m taller than the Sears Tower in Chicago, Illinois, USA. The Skyscrapers (Taipei, Taiwan, 2004) has height of 509 m, the Eiffel Tower (Paris, 1887-1889) has 300 m, Empire State Building (USA, New York, 1930-1931) has 381 m + TV mast of 61 m. Under construction a building of 1001 m (Kuwait City, Kuwait) and 1430 m Supported Structure in Gulf of Mexico.

Current materials make it possible even today to construct towers many kilometers in height. However, conventional towers are very expensive, costing billions of dollars. When considering how high a tower can be built, it is important to remember that it can be built to any height if the base is large enough. Theoretically, you could build a tower to geosynchronous Earth orbit (GEO) out of bubble gum, but the base would likely cover half the surface of the Earth.

The proposed optimal masts (towers) are cheaper in lots of hundreds. They can be built on the Earth's surface and their height can be increased as necessary. Their base is not large. The main innovations in this project are the application of optimal structures (minimum weight), hydrogen for filling tube structures at high altitude and a solution of a stability problem for tall (thin) solid columns, and utilization of new materials [4]-[7].

**The tower applications**. The high towers (3-100 km) have numerous applications for government and commercial purposes:

- Entertainment and Observation platform.
- Entertainment and Observation desk for tourists. Tourists could see over a huge area, including the darkness of space and the curvature of the Earth's horizon.
- Drop tower: tourists could experience several minutes of free-fall time. The drop tower could provide a facility for experiments.
- A permanent observatory on a tall tower would be competitive with airborne and orbital platforms for Earth and space observations.
- Communication boost: A tower tens of kilometers in height near metropolitan areas could provide much higher signal strength than orbital satellites.
- Solar power receivers: Receivers located on tall towers for future space solar power systems would permit use of higher frequency, wireless, power transmission systems (e.g. lasers).
- Low Earth Orbit (LEO) communication satellite replacement: Approximately six to ten 100-km-tall towers could provide the coverage of a LEO satellite constellation with higher power, permanence, and easy upgrade capabilities.

Other new revolutionary methods of access to space are described in [8]-[15].

## 2. Description of Innovation and Problem

2.1. **Tower structure.** The simplest tourist tower includes (fig.1): Solid mast, top observation desk, elevator, expansions, and control stability. The tower is separated into sections by horizontal and vertical rods (fig.2) and contains control devices.

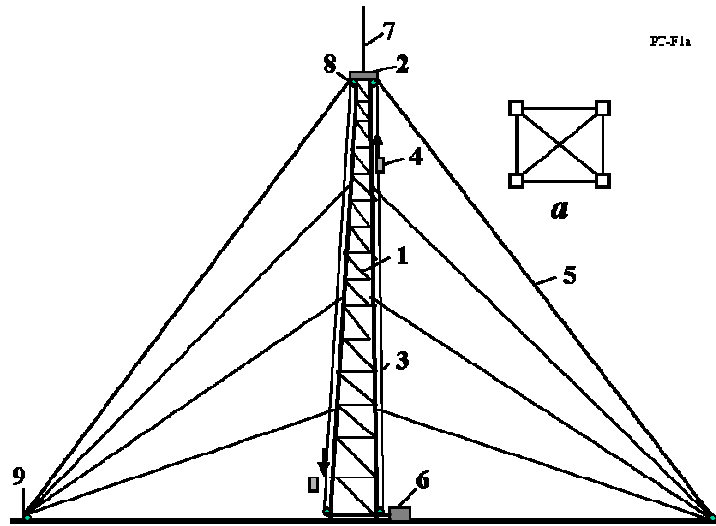

**Fig.1**. Solid optimal space tower (mast) of height 3 - 100 km. (a) typical cross-section of tower. Notations: 1 – solid column; 2 – observation desk; 3 – load cable elevators; 4 – passenger cabin; 5 – expansions; 6 – engine; 7 – radio and TV antenna; 8 – rollers of cable transport system; 9 – stability control.

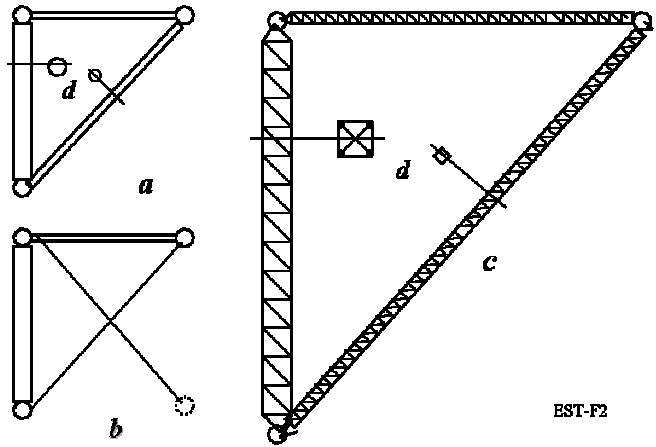

**Fig. 2**. Section of optimal solid tower. Notation: (a) - the first level tube rod with sold diagonal braces; (b) - the first level rod with flexible braces; (c) - the second level rod with lattice column and braces; (d) - cross-section of rods.

2.2. **Filling gas**. The compressed gas should fill the tube tower rods that provide the structure's weight. Author suggests filling the towers with a light gas, for example, hydrogen.

The average temperature of the atmosphere in the interval from 0 to 100 km is about 240°K.

2.3. **The observation radius** versus altitude is presented in [8], figs.4-5 [Eq. (23)].

2.4. **Tower material**. The tower parameters very depend on the strength of material, specifically the relation of the safety press stress $\sigma$ to specific density $\gamma$. Pressure limit is approximately three times more then tensile stress for most conventional materials.

The properties of the some current materials are presented in Table 1.

**Table 1**. Compressive strength of some materials (Kikoin [4], ps. 38, 41, 52, 54)

| Material | Density $\gamma$ [kg/m$^3$] | Pressure limit $10^{-7}\sigma$ [N/m$^2$] | Strength coefficient $K=10^{-7}\sigma/\gamma$ | Tensile stress $10^{-7}\sigma$ [N/m$^2$] |
|---|---|---|---|---|
| Steel, 40X | 7900 | 400 | 0.050 | 120 |
| Alloy WC | 19000 | 600 | 0.032 | 110 |
| Duralumin | 2900 | 150 | 0.052 | 54 |
| Quartz | 2650 | 1200 | 0.453 | - |
| Corundum | 4000 | 2100 | 0.525 | - |
| Diamond | 3520 | 9859 | 2.8 | - |

Current industry widely produces artificial fibers having tensile stress $\sigma = 500 - 620$ kg/mm$^2$ and density $\gamma = 1800$ kg/m$^3$. Their tensile ratio is $K = 10^{-7}\sigma/\gamma = 0.28 - 0.34$. There are whisker (in industry) and nanotubes (in scientific laboratory) having tensile $K = 1 - 2$ (whisker) and $K = 5 - 11$ (nanotubes). Theory predicts fiber, whisker and nanotubes having $K$ ten times greater [5]-[7]. These materials can be used for light guy-lines.

The tower parameters have been computed for pressure $K = 0.05 - 0.3$. Recommend value for guy-lines is $K = 0.1$.

2.5. **Tower safety**. For safety of people (passenger cabin) parachutes can be used.

2.6. **Tower stability**. Stability is provided by expansions (tensile elements). The verticality of the tower (mast) can be checked by laser beam and GPS sensors monitoring beam location (fig.2). If a section deviates from vertical control cables, control devices automatically restore the tower position.

2.7. **Tower construction**. The tower building will not have conventional construction problems such as lifting building material to high altitude. The tower (mast) is not heavy. New sections are put under the tower, the new section is lifted, and the entire tower is lifted. It is estimated the building may be constructed in 4 -12 months. A small tower (up to 3 km) can be located in city.

2.8. **Tower cost.** The tower does not require high-cost building materials. The tower will be a tens times cheaper than conventional reinforced concrete towers 400 - 600 m tall.

## 3. Theory of optimal solid tower

Equations developed and used by author for estimations and computation are provided below.

1. **Optimal cross-section area for solid tower of compressive stress**. Optimal cross-section area for space elevator cable (tensile stress) the author received in [9], Eqs. (1) - (5), (see also [10], Ch.1). For compressive stress we must change the sign (" -") at value *B*. The equation (4) for our case (rotary Earth and variable gravity) is

$$\overline{A}(R) = \frac{A}{A_0} = \exp\left[-\frac{\gamma g_0 B(R)}{\sigma}\right], \quad B(R) = R_0^2\left\{\left(\frac{1}{R_0} - \frac{1}{R}\right) - \frac{\omega^2}{2g_0}\left[\left(\frac{R}{R_0}\right)^2 - 1\right]\right\}, \quad (1)$$

$$\overline{M} = \frac{M}{G} = \frac{g_0}{k\overline{A}(R)}\int_{R_0}^{R}\overline{A}(r)dr, \quad k = \frac{\sigma}{\gamma}, \quad K = 10^{-7}k,$$

where *A* is cross-section area of solid tower, m²; $A_0$ is initial (at ground) cross-section area, m²; $\overline{A}$ is relative cross-section area of tower (mast); *R* is radius (distance from Earth center), m; $R_0$ is Earth radius, m, $R_0$ = 6.378 km; $g_0$ = 9.81 m/s2 is Earth gravity at Earth surface; $\omega$ = 72.685×10$^{-6}$ rad/s is Earth angle speed, *G* is vertical force at tower top, kg; *M* is tower weight, kg; $\overline{M}$ is relative tower weight (weight for every unit load mass).

If the gravity is constant and Earth does not rotate, the equation (1) is simpler

$$\overline{A} = \exp\left[-\frac{\gamma g_0 H}{\sigma}\right] = \exp\left[-\frac{g_0 H}{k}\right], \quad \text{where} \quad k = \frac{\sigma}{\gamma}, \quad \overline{M} = \frac{M}{G} = \left(e^{\frac{gh}{k}} - 1\right). \quad (2)$$

The computations for tower height *H* = 100 km and for tower *H* = 37,000 km (geosynchronous orbit) are presented in fig. 3 - 7.

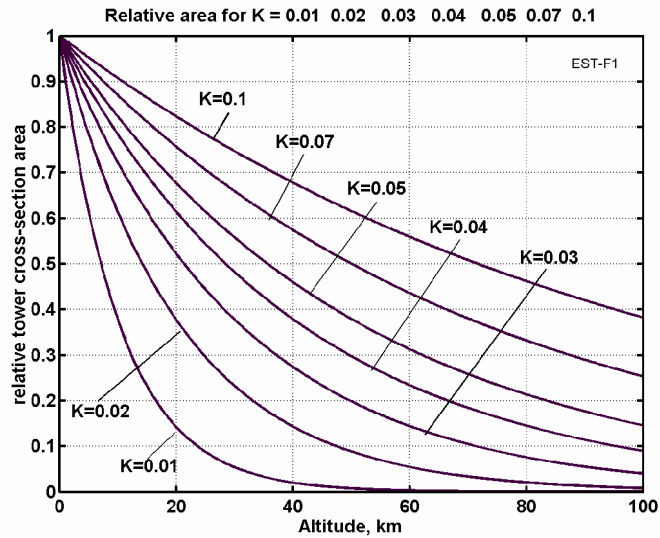

**Fig.3**. Relative tower cross-section aria versus tower altitude (up 100 km) and pressure strong coefficient.

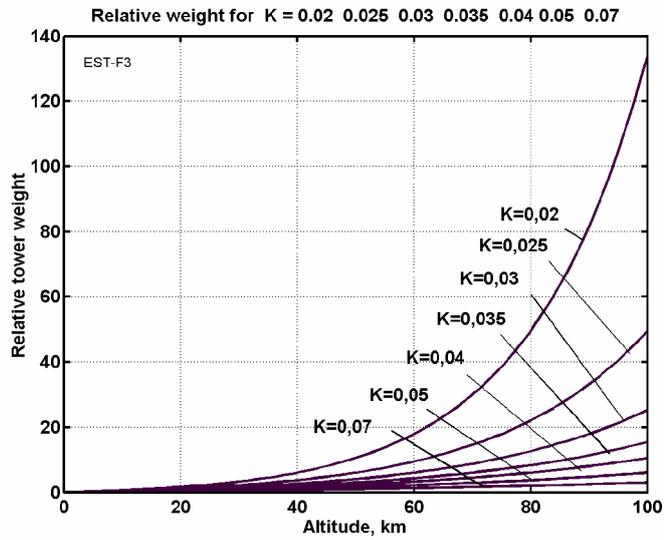

**Fig. 4**. Relative tower mass for height $H$=100 km versus pressure stress coefficient $K$.

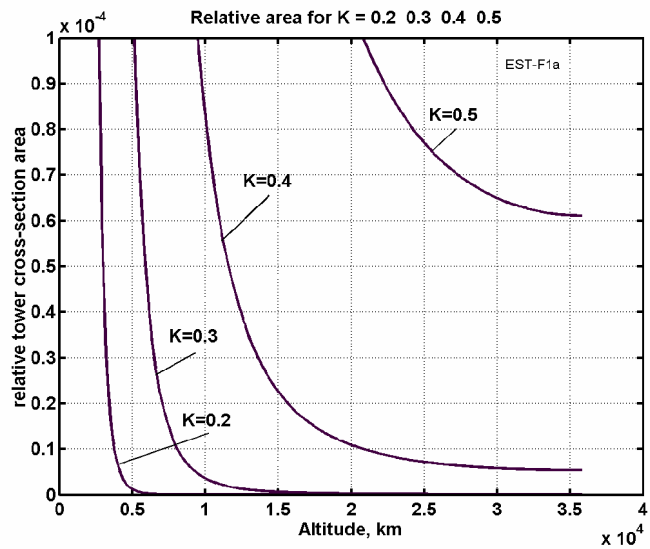

**Fig. 5**. Relative cross-section ratio $S/S_0$ for the tower height $H$=37,000 km (geosynchronous orbit) versus the pressure stress coefficient $K$.

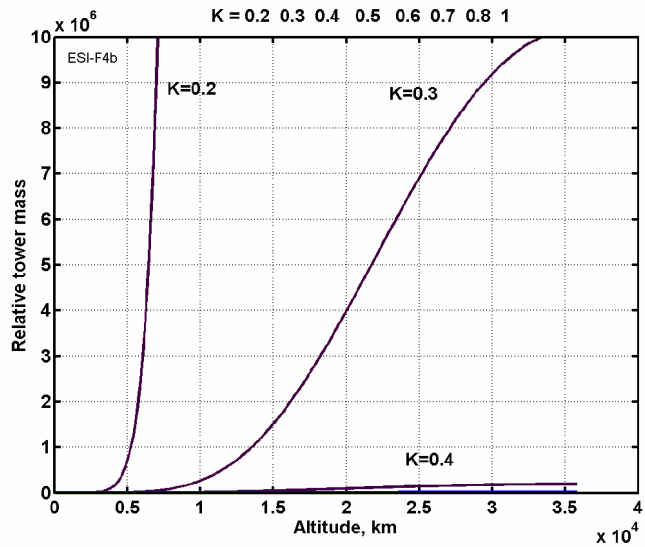

**Fig. 6**. Relative tower mass for tower height $H$=37,000 km (geosynchronous orbit) versus pressure stress coefficient $K$ = 0.2 - 0.4.

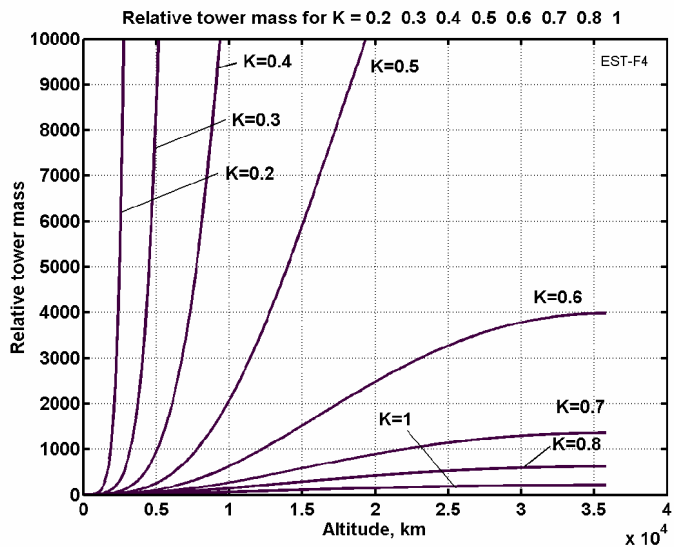

**Fig. 7**. Relative tower mass for tower height $H$=37,000 km (geosynchronous orbit) versus pressure stress coefficient $K$ = 0.2 - 1.

The figs. 3 - 4 show the optimal steel tower (mast) having the height 100 km, safety pressure stress $K$ = 0.02 (158 kg/mm$^2$) must have the bottom cross-section area approximately in 100 times more then top cross-section area and weight is 135 times more then top load (fig. 4). For example, if full top load equals 100 tons (30 tons support extension cable + 70 tons useful load), the total weight of main columns 100 km tower-mast (without extension cable) will be 13,500 tons . It is less that a weight of current sky-scrapers (compare with 3,000,000 tons of Toronto tower having the 553 m height). In reality if the safety stress coefficient $K$ = 0.015, the relative cross-section area and weight will sometimes be more but it is a possibility of current building technology.

The figs. 5 - 7 show the building of the geosynchronous tower-mast (include the optimal tower-mast) is very difficult. For $K$ = 0.3 (it is over the top limit margin of safety for quartz, corundum) the tower mass is ten millions of times more than load (fig. 6), the extensions must be made from nanotubes and they weakly help. The problems of stability and flexibility then appear. The situation is strongly improved if tower-mast built from diamonds (relative tower mass decreases up 100, fig.7). But it is not known when we will receive the cheap artificial diamond in unlimited amount and can create from it building units.

**2. Using the compressive rods** [9]. The rod compressed by gas can keep more compressive force because internal gas makes a tensile stress in a rod material. That longitudinal stress cannot be more then a half safety tensile stress of road material because the compressed gas creates also a tensile radial rod force (stress) which is two times more than longitudinal tensile stress. As the result the rod material has a complex stress (compression in a longitudinal direction and a tensile in the radial direction). Assume these stress is independent. The gas has a weight which must be added to total steel weight. The author used the following equations for computation of the gas compressive rods

$$\sigma_g = \sigma_c + \frac{1}{2}\sigma_t, \quad \gamma_g = \gamma_0 + \frac{rp}{2\delta} = \gamma_0 + \frac{\mu\sigma_t}{2RT}, \quad K_g = \frac{\sigma_g}{\gamma_g}, \quad p = \frac{\rho RT}{\mu}, \tag{3}$$

where $\sigma_g$ is safety stress of gas compressed rod, N/m$^2$; $\sigma_c$ is safety load compressed stress, N/m$^2$; $\sigma_t$ is safety tensile gas stress, N/m$^2$; $\gamma_g$ is specific density of gas compressed rod, kg/m$^3$; $\gamma_0$ is specific density conventional rod, kg/m$^3$; $\mu$ is the gas molar weight (for hydrogen $H_2$ it equals $\mu = 0.002$ kg/mole), $R = 8.314$ is constant, $T$ is temperature), $^\circ$K; $p$ is gas pressure, N/m$^2$; $\rho$ is gas density, kg/m$^3$; $\delta$ is wall thickness of rod, m; $r$ is rod radius, m.

For steel and duralumin from Table 1, the internal gas increases $K$ in 35 - 45%.

Unfortunately, the gas support depends on temperature (see Eq. (3)). That means the mast can loss this support at night. Moreover, the construction will contain the thousands of rods and some of them may be not enough leakproof or lose the gas during of a design lifetime. I think it is a danger to use the gas pressure rods in space tower.

## Conclusion

The inexpensive steel tower-mast of the height up 100 - 200 km (and more) can be built without big problems at the present time. They can be useful for communication (TV, radio, telephone), for radiolocation (defense), for space launch, for tourism (include space tourism), for scientists (astronomy), for solar energy, and for many other applications. The offered optimal design allows finding of the minimum of a tower-mast weight which can be reached in this space building.
The other designs of space towers are in [8]-[15].

## Acknowledgement

The author wishes to acknowledge Richard Cathcart for correcting the English and useful advices.